\documentclass{iau} 
\usepackage{graphicx}

\title[IAU Symp. 322~~Astrochemistry VII] 
{Fire from Ice - Massive Star Birth from Infrared Dark Clouds}

\author[Jonathan C. Tan]   
{Jonathan C. Tan$^{1,2,3}$}

\affiliation{
$^1$Dept. of Space, Earth and Environment, Chalmers University, SE-412 96 Gothenburg, Sweden\\
$^2$Dept. of Astronomy, University of Virginia, Charlottesville, VA 22904, USA\\ 
$^3$Depts. of Astronomy and Physics, University of Florida, Gainesville, FL 32611, USA\\
email: {\tt jctan.astro@google.com}}

\pubyear{2017}
\volume{322}  
\jname{Astrochemistry VII}
\editors{M. Cunningham, T. Millar \& Y. Aikawa, eds.}
\begin{document}

\maketitle

\begin{abstract}
I review massive star formation in our Galaxy, focusing on initial
conditions in Infrared Dark Clouds (IRDCs), including the search for
massive pre-stellar cores (PSCs), and modeling of later stages of
massive protostars, i.e., hot molecular cores (HMCs). I highlight how
developments in astrochemistry, coupled with rapidly improving
theoretical/computational and observational capabilities are helping
to improve our understanding of the complex process of massive star
formation.
\end{abstract}

\firstsection 
\section{Introduction}

Massive stars and their associated star clusters are important
throughout astrophysics, yet there remain many open questions about
how they form. These include: What is the accretion mechanism of
massive star formation?  What sets the initial mass function of stars,
especially at the highest masses? What is the relation of massive star
formation to star cluster formation? How do massive star and star
cluster formation vary with galactic environment? The nature of the
very first, metal-free, i.e., Population III, stars, often theorized
to have been of high-mass, is an enduring open question.

In this short review, I focus on theoretical models and observational
studies of local, i.e., Galactic, massive star and star cluster
formation, concentrating on developments since the reviews of Tan et
al. (2014, hereafter T14) and Tan (2016) and offering a somewhat
different perspective compared to the review of Motte et al. (2017). I
emphasize the importance of incorporating astrochemistry into
theoretical models that try to predict the physical outcome of the
massive star formation process, especially to predict the ionization
fraction that couples the mainly neutral gas to magnetic
fields. Astrochemistry is also crucial for providing diagnostics
needed to test different theoretical models.

Following Williams et al. (2000) and T14, we define gas {\it clumps}
as self-gravitating structures that fragment into star clusters,
perhaps via a population of self-gravitating pre-stellar {\it cores}
(PSCs). These cores are defined to be structures that collapse to a
central, rotationally-supported disk leading to single or small-$N$
multiple star formation. As massive stars tend to form in clusters
(e.g., de Wit et al. 2005), massive star formation and cluster
formation are connected processes that need to be understood together.

The observed range of masses of star-forming clumps, including early
stage Infrared Dark Clouds (IRDCs), and young star clusters is from
$M\sim 10$ to $\sim10^6\:M_\odot$ (including the most extreme ``super
star clusters'' seen in some relatively nearby galaxies, such as M82,
NGC 1569 and NGC 5253), while mass surface densities, $\Sigma$,
typically range from $\sim0.03$ to $\sim 10\:{\rm g\: cm}^{-2}$ (i.e.,
$\sim200$ to $\sim 5\times 10^4\:M_\odot\: {\rm pc}^{-2}$) (T14). Thus,
protocluster clumps have radii of $\sim1$ to 10~pc and average H
nuclei number densities of $n_{\rm H}\sim10^3$ to $\sim10^{6}\:{\rm
  cm}^{-3}$, although they may contain higher density substructures,
including cores.

\section{Theoretical Models of Massive Star Formation}


\subsection{Overview of Formation Scenarios}\label{S:scenarios}

There is a long-standing debate about the basic formation mechanism of
massive stars. Theories range from Core Accretion models (e.g.,
McLaughlin \& Pudritz 1997; McKee \& Tan 2003 [hereafter MT03], who
presented the Turbulent Core Model [TCM]) that are scaled-up versions
of standard low-mass star formation theories (Shu, Adams \& Lizano
1987), to Competitive Accretion models at the crowded centers of
forming star clusters (Bonnell et al. 2001; Wang et al. 2010) that do
not involve massive starless cores, to Protostellar Collisions
(Bonnell et al. 1998; Bally \& Zinnecker 2005; Moeckel \& Clarke
2011). Still, as Core Accretion models are the most conservative
and simplest way to treat massive star formation and potentially offer
a universal mechanism by which to understand all star formation, we
will focus mostly on their predictions and comparison to observations,
noting cases where these seem to be a relatively poor description.

\subsection{Initial Conditions of the Turbulent Core Model: Massive Pre-Stellar Cores}

The TCM assumes the initial conditions of massive stars are massive
PSCs, approximated as polytropic spheres, that are in quasi virial
equilibrium and pressure equilibrium with their surrounding clump
environment. This implies that a core of mass $M_c$ in a
self-gravitating clump of mass surface density $\Sigma_{\rm cl}$,
which sets ambient pressure, has radius
\begin{equation}
R_c \simeq 0.0574 (M_c/60\:M_\odot)^{1/2}(\Sigma_{\rm
    cl}/1\:{\rm g\:cm}^{-2})^{-1/2}\:{\rm pc},
\end{equation}
(MT03), i.e., 12,000~AU, for fiducial parameter choices, including
index of internal power law density profile, $k_\rho=1.5$. The core
has a mass-averaged 1D velocity dispersion of
\begin{equation}
\sigma_c \simeq 1.09 (\phi_{B}/2.8)^{-1/2} (M_c/60\:M_\odot)^{1/4}(\Sigma_{\rm cl}/1\:{\rm g\:cm}^{-2})^{1/4}\:{\rm km\:s}^{-1},
\label{eq:sigma}
\end{equation}
where $\phi_{B}=1.3+1.5 m_A^{-2}$ accounts for the effects of
$B$-fields, taking a value of 2.8 in the fiducial case of an Alfv\'en
Mach number, $m_A$, of unity. However, we note that MT03 approximated
the magnetic pressure $B^2/(8 \pi)$ as isotropic, which it is not. A
random magnetic field has an isotropic pressure of $B^2/(24 \pi)$,
which would reduce $\phi_B$, i.e., in this limit the fiducial value
would become about 1.8 (C. McKee, private communication).

Note, these conditions apply to the gas structure at the moment just
before protostar formation and little is yet assumed about how the
core itself formed. By definition, it will have emerged from the clump
gas, perhaps via top-down fragmentation, i.e., a massive starless core
condensing out of the ambient clump medium, or via bottom-up growth of
a smaller gravitationally bound starless core that gains mass by some
combination of accretion from the clump or mergers with other cores
(such core merging should not be confused with protostellar
mergers). The timescale over which the core forms is also not
specified: a range of possibilities from very fast, i.e., $t_{\rm
  c,form}\sim 1 t_{\rm ff}$, where $t_{\rm
  ff}=(3\pi/[32G\rho])^{1/2}=1.4\times10^5(n_{\rm H}/10^5\:{\rm
  cm}^{-3})^{-1/2}\:$yr is the local free-fall time of the core given
its mean density $\rho$ or equivalently mean H nuclei number density,
$n_{\rm H}$, to very slow, i.e., $t_{\rm c,form}\gtrsim10 t_{\rm ff}$
may be considered. If core formation is a relatively slow process,
i.e., $\gtrsim$ a few $t_{\rm ff}$, then the conditions of quasi
virial and pressure equilibrium are more likely to be achieved. Note
also that while the boundary of a PSC has a precise theoretical
definition, i.e., delimiting the material that is gravitationally
bound to the core, observationally it can be very challenging to
isolate core material by this criterion, especially since the mass
surface densities of the core and the surrounding clump have similar
values (MT03). We discuss observational definitions of PSCs in
\S\ref{S:observation_PSC}.

The mass scale of a core may potentially be set by its degree of
magnetization. If it forms relatively slowly from a more strongly
magnetized state as magnetic flux is gradually removed from the region
via ambipolar (e.g., 
Mouschovias 1991) or reconnection (Lazarian \& Vishniac 1999; Eyink et
al. 2011) diffusion, then it will be close to the boundary between a
sub- and super-critical state, i.e., where the $B$-field strength is
just strong enough to impede collapse. Then the mass of the
contracting core is effectively set by the magnetic critical mass
(e.g., Bertoldi \& McKee 1992)
\begin{equation}
M_{\rm c,B} = 51.8 (y/0.5)^{-2} (B_c/200\:{\rm \mu G})^3 (n_{\rm H,c}/10^5\:{\rm cm}^{-3})^{-2}\:M_\odot,
\label{eq:McB}
\end{equation}
where $y\equiv Z/R$ is the aspect ratio ($R$ being the radius normal
to axis of symmetry for an ellipsoidal core; $2Z$ being size of core
along this axis) here normalized to a moderate degree of flattening
along the field direction, $B_c$ is the average magnetic flux in the
core (here normalized to a typical value inferred from observations of
molecular clouds at this density, Crutcher et al. 2010) and $n_{\rm
  H,c}$ is the mean density of the core. Kunz \& Mouschovias (2009)
have argued that the entire PSC mass function (PSCMF) may be set by
modest variations in the degree of magnetization of gas within
star-forming regions.

Eq.~(\ref{eq:sigma}) implies massive PSCs are expected to have
internal turbulent motions much greater than the isothermal sound
speed at $\sim10\:$K, i.e., $c_{\rm th}=0.19 (T/10\:{\rm
  K})^{1/2}\:{\rm km\:s}^{-1}$. Thus they are likely be supersonically
turbulent, which would induce internal sub-structure. The resulting
localized shock heating, i.e., up to $\sim50\:$K for $\sim1\:{\rm
  km\:s}^{-1}$ shock speeds, is expected to have astrochemical
effects, e.g., liberation of species from grain ice mantles if their
temperatures are well-coupled to the gas (CO freeze-out being
efficient at $\lesssim 20\:$K) and reduction in the level of
deuteration of key diagnostic species $\rm N_2D^+$ (Kong et al. 2015,
hereafter K15, see below). Dissipation of energy in these shocks leads
to a reduction in the level of turbulence with decay times of about a
few $t_{\rm ff}$ (e.g., McKee \& Ostriker 2007), which may be
partially offset by gravitational contraction of the core. If
diffusion of $B$-field out of the core occurs more slowly, then one
expects an evolution towards a state with a lower degree of
turbulence, i.e., smaller values of $m_A$, and a stronger degree of
support from large-scale $B$-fields.

Deuterated species, such as $\rm N_2D^+$, are thought to be one of the
key observational tracers of PSCs (\S\ref{S:observation_PSC}). This
expectation is based on studies of nearby, relatively low-mass PSCs,
such as L1544 (e.g., Caselli \& Ceccarelli 2012), with observations of
such sources leading to the refinement of astrochemical models that
aim to predict the abundance of these diagnostic species. High levels
of deuteration arise
because in the cold, dense conditions relevant to PSCs, CO molecules
are mostly frozen-out onto dust grain ice mantles, which allows the
abundance of $\rm H_2D^+$ to build up via the exothermic reaction of
$\rm H_3^+ + HD \rightarrow H_2D^+ + {\rm p-}H_2$, where we have
indicated this applies only for the para form of $\rm H_2$. The ortho
form of $\rm H_2$ has enough energy to drive the reverse reaction. As
the ortho-to-para ratio of $\rm H_2$ ($\rm OPR^{H2}$) drops, e.g.,
mediated by either gas phase (e.g., Sipil\"a et al. 2015; K15) or
grain surface phase (Bovino et al. 2017) reactions, this then leads to
significant ``deuteration fractions'', $D_{\rm frac}$, i.e., abundance
ratio to the non-deuterated species, of $\rm H_2D^+$ and $\rm N_2D^+$
(the latter formed via $\rm H_2D^++N_2\rightarrow H_2 + N_2D^+$).

The gas-phase astrochemistry associated with this deuteration, i.e.,
including ortho and para spin states, has been studied in the context
of one-zone models by, e.g., Sipil\"a et al. (2010, 2015), Wirstr\"om
et al. (2012) and K15, with the latter presenting an exploration of
the parameter space relevant to IRDCs/protoclusters and massive PSCs
(i.e., $n_{\rm H}=10^3$ to $10^7\:{\rm cm}^{-3}$, $T=10$ to $30\:$K,
cosmic ray ionization rates $\zeta=10^{-18}$ to $10^{-15}\:{\rm
  s}^{-1}$, assumed fixed gas-phase heavy element depletion factors (a
proxy for CO freeze-out) from $f_D=1$ (no depletion) to $1000$ (i.e.,
total abundance divided by gas phase abundance is 1000), and initial
values of $\rm OPR^{\rm H2}$ from $\sim10^{-3}$ to 3. Some of the main
results of this study were the near-equilibrium values of $D_{\rm
  frac}^{\rm N_2D^+}$ under given conditions and the timescales needed
to reach 90\% of these near-equilibrium conditions. For example, for
conditions of the fiducial TCM, i.e., $n_{\rm H}\simeq10^5\:{\rm
  cm}^{-3}$ and $T=15\:$K (and $\zeta=2.5\times10^{-17}$, $f_D=10$),
K15 found $D_{\rm frac,eq}^{\rm N_2D^+}=0.181$ and the time to reach
within 90\% of this value to be $t_{\rm eq,90}=1.25$~Myr (for initial
$\rm OPR^{\rm H2}=0.1$). The fact that this is almost ten times longer
than the local free-fall timescale for this density is the reason why
the deuteration process may be a useful ``chemical clock'' to
constrain the dynamics of core formation. However, this does depend on
correct modeling of the timescale of conversion of ortho to para $\rm
H_2$ (see Bovino et al. 2017), as well as the assumption made for the
``initial'' value of $\rm OPR^{H2}$. Observational constraints on $\rm
OPR^{H2}$ are important for breaking such model degeneracies
(\S\ref{S:observation_PSC}).

\begin{figure}[t]
\vspace{-0.1 cm}
\begin{center}
\includegraphics[width=5.3in]{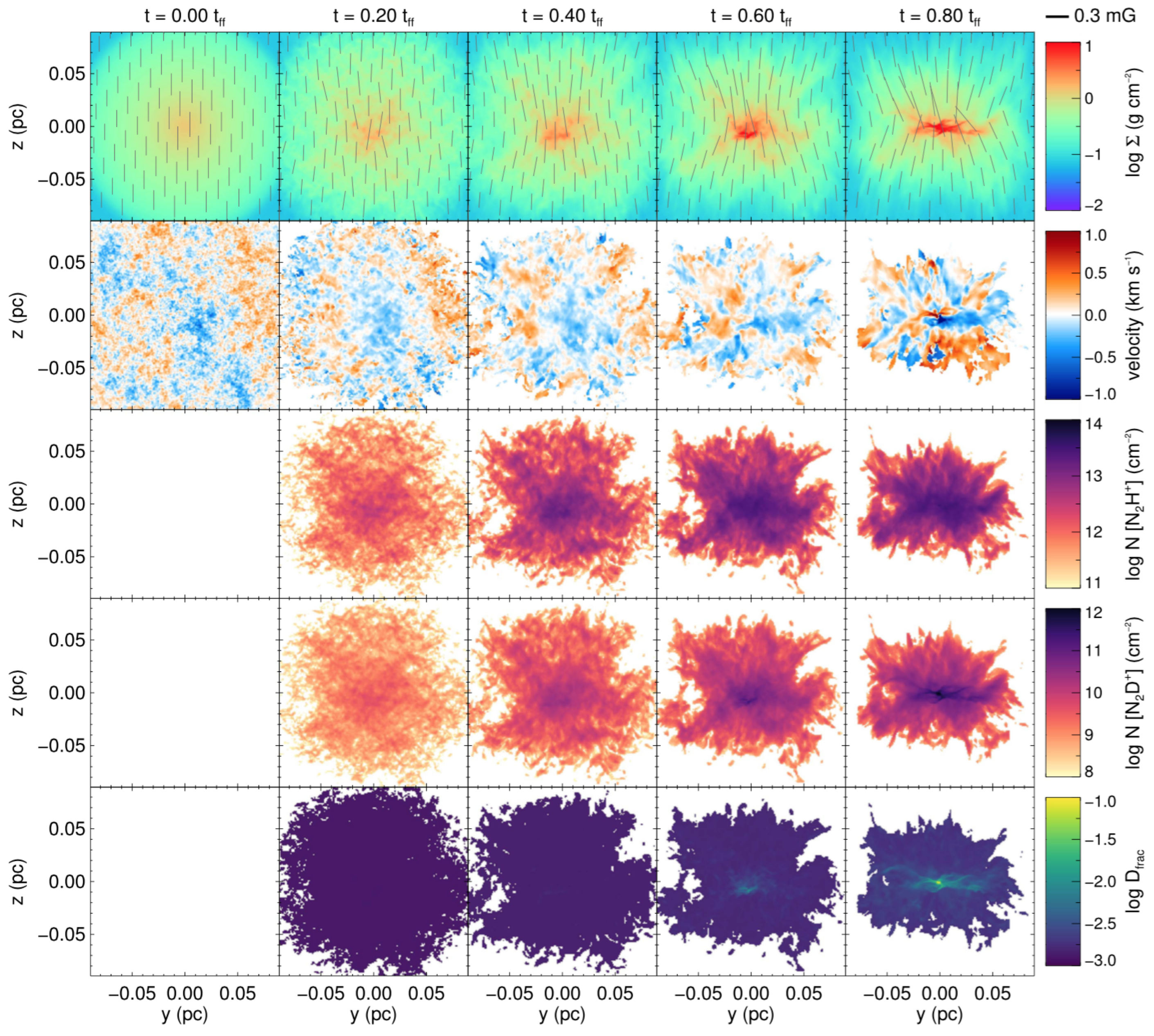} 
\vspace{-0.6cm}
 \caption{
%
%
Evolution of a $60\:M_\odot$ PSC embedded in $\Sigma=0.3\:{\rm g\:
  cm^{-2}}$ clump (Goodson et al. 2016) with properties set by the TCM
(MT03). Time evolution is left to right from 0 to 0.8~$t_{\rm ff}$,
with initial mean local free-fall time, $t_{\rm ff} = 0.2\:$Myr. 1st
row: mass surface density, $\Sigma$. Total projected $B$-field
strengths and orientations are shown by lines (scale in top
right). 2nd row: mean velocity of $\rm N_2D^+$. 3rd and 4th rows: $\rm
N_2H^+$ and $\rm N_2D^+$ column densities, via analytic fits to the
K15 model. 5th row: $D_{\rm frac}^{\rm N2H+}\equiv [{\rm
    N_2D^+}]/[{\rm N_2H^+}]$. This particular simulated core undergoes
relatively fast collapse, so there is little time to build-up very
high values of $D_{\rm frac}^{\rm N2H+}$.}
\label{fig:goodson}
\end{center}
\end{figure}

K15 also presented a series of chemodynamical models, including
evolution of $f_D$ by freeze-out and evolution of $n_{\rm H}$ at rates
relative to that of free-fall collapse, i.e., following $dn_{\rm H}/dt
= \alpha_{\rm ff}n_{\rm H}/t_{\rm ff}$. Again, since the time to reach
high levels of deuteration of $\rm N_2D^+$ can be significantly longer
than a local free-fall time, these models can be used in conjunction
with observed values of $D_{\rm frac}^{\rm N_2D^+}$ to constrain
$\alpha_{\rm ff}$ (\S\ref{S:observation_PSC}). Such chemodynamical
models can be improved to include core radial structure; see, e.g.,
Pagani et al. (2009) for an example of a simple astrochemical model
(no spin-state chemistry or N chemistry) applied to low-mass cores and
the more general modeling of Gerner et al. (2015). Full 3D numerical
(M)HD simulation that is fully coupled to an astrochemical network
following deuteration is challenging.
For example, K\"ortgen et al. (2017) carried out simulations with a
reduced network that focused on $\rm H_2D^+$ ($\rm N_2D^+$ was not
modeled). Taking an intermediate approach, Goodson et al. (2016)
analyzed the K15 models to derive analytic expressions for the growth
rate of $\rm N_2H^+$ and $\rm N_2D^+$ abundances as a function of
local density and starting $\rm OPR^{H2}$. These expressions were
incorporated in look-up tables to estimate abundances of these species
in MHD simulations of massive cores (Fig.~\ref{fig:goodson}). These
results can be compared to observed candidate massive PSCs
(\S\ref{S:observation_PSC}).

In addition to PSC diagnostics, astrochemical modeling is also needed
to estimate ionization fractions in these clouds. For example, for the
fiducial parameters of the K15 gas phase astrochemical modeling
described above, in particular with $n_{\rm H}=10^5\:{\rm cm}^{-3}$
and $\zeta=2.5\times10^{-17}\:{\rm s}^{-1}$, $n_e/n_{\rm H}$ is found
to range from $\sim10^{-8}$ when $f_D=1$ (no depletion) up to $\sim
5\times 10^{-8}$ in the limit of high $f_D$. In the case of $f_D=10$,
the positive charge carriers are dominated by about equal fractions of
$\rm H_3^+$ and $\rm HCO^+$. The ambipolar diffusion time, $t_{\rm
  ad}=2.5\times10^{13}(n_e/n_{\rm H})\:{\rm yr}$, then has a fiducial
value of $5\times10^5\:$yr, i.e., about three times greater than
$t_{\rm ff}$. As discussed below, the ionization fraction also depends
on assumptions about the dust grain size distribution, which
introduces further uncertainties.

\subsection{Massive Protostellar Cores}

In Core Accretion models, a very low-mass ($\ll1\:M_\odot$) protostar
is first expected to form near core center. This acts as a seed for
continued accretion from the infall envelope and the structure is now
known as a protostellar core. In the fiducial TCM with collapse at
rate similar to that of local free-fall, the accretion rate delivered
by the infall envelope is
\begin{equation}
\dot{m}_{*d} = 9.3\times 10^{-4} \epsilon_{*d} (M_{c}/60\:M_\odot)^{3/4}(\Sigma_{\rm cl}/1\:{\rm g\:cm}^{-2})^{3/4}(M_{*d}/M_c)^{1/2}\:M_\odot\:{\rm yr}^{-1},
\label{eq:mdot}
\end{equation}
where $\dot{m}_{*d}$ is the rate of increase of the mass of the
protostar and its disk, $\epsilon_{*d}$ is the current efficiency of
the infall rate with respect to uninhibited collapse (values $\sim
0.5$ are expected to develop due to protostellar outflow feedback,
e.g., Zhang, Tan \& Hosokawa 2014 [ZTH14]) and $M_{*d}$ is the
idealized collapsed mass supplied to the central disk in the
no-feedback limit. Here the ratio $M_{*d}/M_c$ indicates the
evolutionary stage of the collapse, i.e., advancing from 0 to 1. The
increasing accretion rates during the collapse are a consequence of
the assumed initial power law density profile of the PSC of
$k_\rho=1.5$. If an index of 2 is adopted, i.e., that of a singular
isothermal sphere, then the accretion rate would be independent of
$M_{*d}/M_c$. With the fiducial rate of eq.~(\ref{eq:mdot}), the star
formation timescale is $t_{*f} = 1.3\times 10^{5}
(M_{c}/60\:M_\odot)^{1/4} (\Sigma_{\rm cl}/1\:{\rm
  g\:cm}^{-2})^{-3/4}\:{\rm yr}$, which has a weak dependence on
$M_c$. Note, $t_{*f}$ is similar to the clump mean free-fall time,
as $t_{*f}/\bar{t}_{\rm
  ff,cl}\rightarrow0.98(M_c/60\:M_\odot)^{1/4}(M_{\rm
  cl}/4000\:M_\odot)^{-1/4}$. If cluster formation takes many
clump free-fall times to complete, then this model allows stars of all
masses to form contemporaneously, i.e., with individual formation
times that are much shorter than that of the cluster.



Initially protostellar accretion may be quasi spherical, i.e.,
directly onto the surface of the protostar that is expected to have a
radius of several $R_\odot$. Later, as material continues to accrete
it is more likely to hit a centrifugal barrier and form an accretion
disk, which then mediates accretion and angular momentum transfer,
including by launching magneto-centrifugal disk winds and X-winds. For
a review of these processes in the context of low-mass star formation,
see, e.g., Inutsuka (2012). However, disk formation is a highly
uncertain process given the expected role of magnetic braking for
transferring angular momentum in the infall envelope (Li et
al. 2014). Indeed, to fully model this process appears to require
following non-ideal MHD processes, such as ambipolar or reconnection
diffusion. To treat ambipolar diffusion requires following the
astrochemical evolution to model the small residual ionization
fraction in the gas that helps mediate coupling of $B$-fields to the
mostly neutral gas. This requires modeling the propagation of cosmic
rays into the central, dense regions of protostellar cores. Zhao et
al. (2016) have also argued that following the abundance of small dust
grains, including PAH molecules, that can act as charge carriers is
also needed to make accurate predictions for disk formation. Given the
uncertain nature of the initial conditions of massive star formation,
e.g., the degree of magnetization of massive PSCs, and given the above
complications of following the coupled chemistry and dynamics of the
collapse, it is not yet possible to make accurate predictions for the
sizes of disks around massive protostars.


The degree of fragmentation in the core will also depend on the
magnetization of the PSC. For core formation mediated by magnetic
fields, little fragmentation is expected below the magnetic critical
mass scale given by eq.~(\ref{eq:McB}). This expectation is broadly
confirmed by the results of numerical simulations. For example, Peters
et al. (2011) simulated a core with $M_c=100\:M_\odot$, $R_c=0.5\:$pc,
$n_{\rm H,c}=5,400\:{\rm cm}^{-3}$ and a weak magnetic field of
$B_c=10\:{\rm \mu G}$. The result of the collapse was a small cluster
of protostars.
Seifried et al. (2011) simulated a core with $M_c=100\:M_\odot$,
$R_c=0.25\:$pc, $n_{\rm H,c}=4.4\times10^4\:{\rm cm}^{-3}$ and
$B_c\simeq1\:{\rm m G}$, while Myers et al. (2013) followed a core
with $M_c=300\:M_\odot$, $R_c=0.1\:$pc, $n_{\rm
  H,c}=2.4\times10^6\:{\rm cm}^{-3}$ and $B_c\gtrsim1\:{\rm m G}$. In
these latter two simulations no fragmentation was seen, with collapse
proceeding approximately monolithically towards a single, central
protostar. The effects of radiative heating (e.g., Krumholz et
al. 2007; Myers et al. 2013) further suppress fragmentation,
especially during the later, higher-luminosity stages.

Semi-analytic treatments of protostar formation and evolution via the
TCM in the limit of no fragmentation have been presented by Zhang \&
Tan (2011), Zhang, Tan \& McKee (2013) and ZTH14. These models include
treatments of the slowly rotating infall envelope, a viscous accretion
disk, disk wind outflows and protostellar evolution. With the density
structure specified by this modeling, the temperature structure is
then calculated via Monte Carlo radiative transfer (RT) simulations,
including both gas and dust opacities, along with emergent radiation,
i.e., to predict multiwavelength images and spectral energy
distributions (SEDs). Zhang \& Tan (2017) present a suite of such
models that cover the parameter space of $M_c=10$ to $480\:M_\odot$
and $\Sigma_{\rm cl}=0.1$ to $3\:{\rm g\:cm}^{-2}$. These models have
been applied to observed sources by De Buizer et al. (2017)
(\S\ref{S:obs_HMC}).


The above models predict that the massive protostar will reach a
H-burning phase while still accreting. The photosphere will have a
high temperature and be a strong source of FUV and EUV radiation,
creating photodissociation regions (PDRs) and HII regions,
respectively. Ionizing feedback will first interact with the
protostellar outflow, creating an ``outflow-confined'' HII region that
will appear as elongated cm continuum emission (Tan \& McKee 2003;
Tanaka et al. 2016). A combination of radiative and mechanical
(outflow) feedback is expected to eventually set the star formation
efficiency from the core, $\epsilon_c$. For $M_c=60\:M_\odot$ and
$\Sigma_{\rm cl}=1\:{\rm g\:cm}^{-2}$, the models of Tanaka et
al. (2017) have $\epsilon_c\simeq0.4$. They also find $\epsilon_c$
decreases for larger $M_c$, as feedback from more massive
stars is more powerful, and decreases as $\Sigma_{\rm cl}$
decreases, since lower density cores, which have lower accretion rates,
are more easily disrupted. Axisymmetric, 2D numerical simulations of
mechanical and radiative feedback from massive protostars have been
presented by Kuiper et al. (2015, 2016) (see also the 3D radiative
feedback only simulations of, e.g., Rosen et al. 2016 and Harries et
al. 2017). One key property that is predicted by the above models and
simulations is the opening angle of the outflow cavity at a given
stage of the protostellar evolution, since this can be compared
directly with observed systems.

The modeling of the astrochemistry of the ``hot molecular cores''
(HMCs) of high-mass protostars has been carried out by a number of
groups (e.g., Charnley et al. 1992; Caselli et al. 1993; Viti et
al. 2004; Doty et al. 2006; Garrod \& Herbst 2006; Aikawa et al. 2012;
\"Oberg et al. 2013; Gerner et al. 2014, 2015). Such modeling, typically
done as post processing of a given physical model, requires specifying
initial chemical conditions, e.g., of the PSC just before star
formation, and then the evolution of densities and temperatures along
streamlines.

\begin{figure}[t]
\vspace{-0.35cm}
\begin{center}
\includegraphics[width=5.3in]{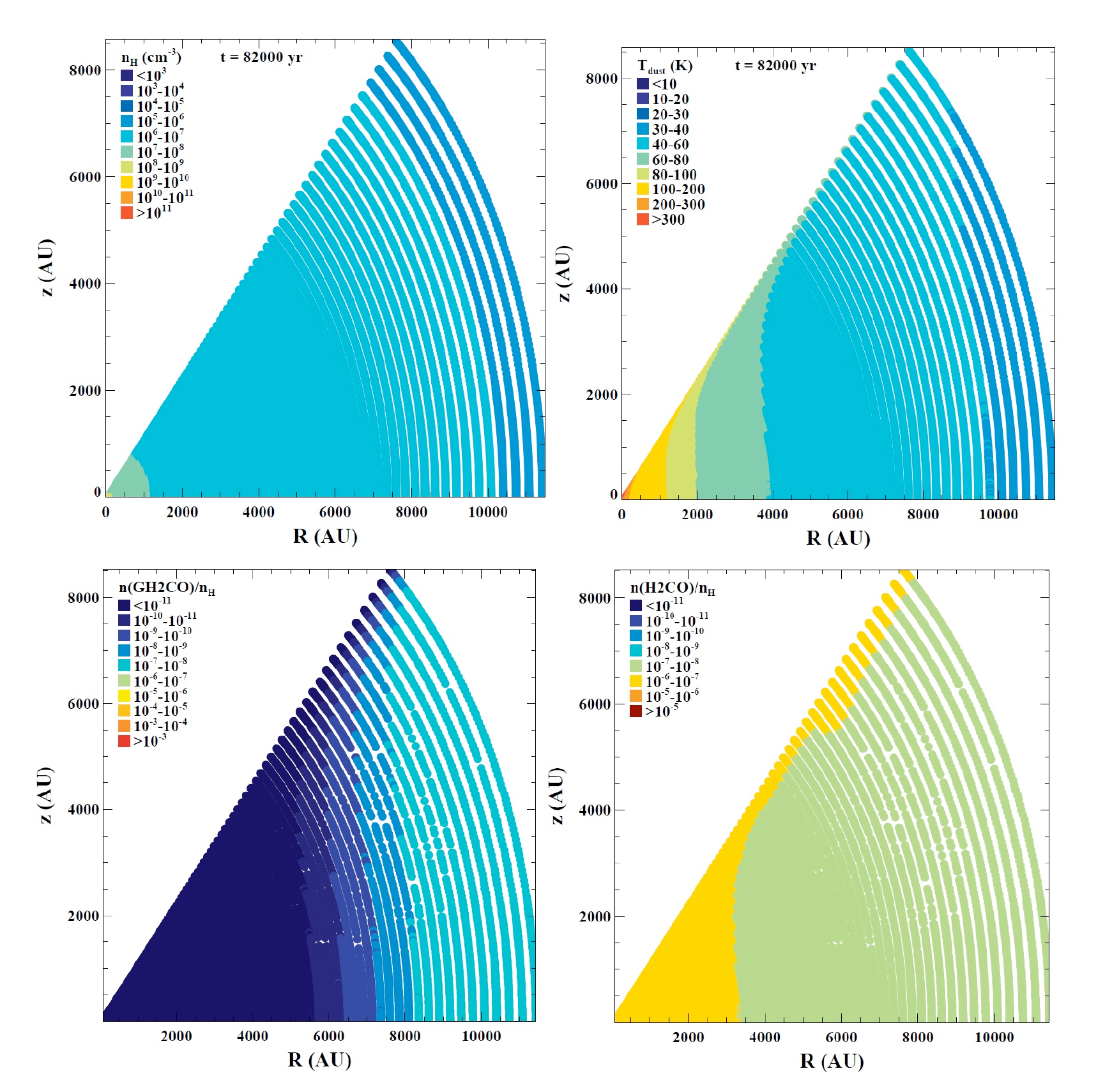} 
\vspace{-0.8cm}
\caption{
(From Drozdovskaya et al., in prep.; see also Drozdovskaya et al. 2017) 
Astrochemical modeling of the infall envelope of a massive protostar
for the case of a $60\:M_\odot$ core embedded in a $\Sigma_{\rm cl} =
1\:{\rm g\: cm}^{-2}$ clump with the protostar now at a mass of
$16\:M_\odot$, located in lower left corner of each panel (ZTH14). By
this point the outflow cavity has opened up to an angle of
$\sim40^\circ$ (white sector: its density, temperature and chemical
structure is not shown here). Note, the accretion disk is also not
modeled here. {\it (a) Top Left:} Density structure ($n_{\rm H}$) at
sampling points along streamlines of the infall envelope. {\it (b) Top
  right:} Temperature structure. {\it (c) Bottom Left:} Gas phase
abundance of $\rm H_2CO$ relative to $n_{\rm H}$. {\it (d) Bottom
  right:} Ice phase abundance of $\rm H_2CO$ relative to $n_{\rm
  H}$. Note, $\rm H_2CO$ starts evaporating from the grain mantles
inside $\sim7,000\:$AU from the protostar and becomes abundant in the
gas phase inside $\sim3,000\:$AU.}
\label{fig:astrochem}
\end{center}
\end{figure}

Figure~\ref{fig:astrochem}, from Drozdovskaya et al. (in prep.), shows
an example of such modeling for a fiducial massive protostellar infall
envelope of the TCM (ZTH14), utilizing a gas-grain network with
$\sim$700 species and 9,000 reactions, including surface chemistry and
complex organic species (Drozdovskaya et al. 2014, 2015). After a
static PSC phase of $3\times×10^5\:$yr, the streamlines of the infall
envelope are followed for $8.2\times10^4\:$yr, i.e., the time needed
to form a $16\:M_\odot$ protostar. The protostar heats the infall
envelope, driving astrochemical reactions in the gas and on grain
surfaces, and liberating ice species. The figure shows examples of
$\rm H_2CO$ gas and ice phase abundances, with the gas phase abundance
rising strongly inside $\sim3,000\:$AU. Such models, with outputs then
coupled to line radiative transfer codes, are important for making
predictions that can be tested with sub-mm/mm observations. However,
the full problem of coupled dynamical and chemical evolution,
including non-ideal MHD effects and the interaction, via shocks, of
protostellar outflows with the infalling core, is challenging to model
accurately, given the many inherent uncertainties in both physical and
chemical aspects.

\subsection{Massive Star Formation in the Context of Star Cluster Formation}

Aspects of star cluster formation have an impact on
massive star formation. As with core formation times, similar
uncertainty applies for clumps, i.e., their formation time from
surrounding (giant) molecular cloud ([G]MC) gas, $t_{\rm cl,form}$,
and the timescale to complete star cluster formation, $t_{\rm
  *cl,form} = (\epsilon/\epsilon_{\rm ff})\bar{t}_{\rm ff,cl}$, where
$\epsilon$ is the final star formation efficiency and $\epsilon_{\rm
  ff}$ is the efficiency per free-fall time. Fast cluster formation in
$\sim 1\bar{t}_{\rm ff,cl}$ has been proposed by Elmegreen (2000, 2007),
Hartmann \& Burkhart (2007) and Hartmann et al. (2012). Slower,
quasi-equilibrium star cluster formation has been proposed by Tan,
Krumholz \& McKee (2006) and Nakamura \& Li (2007, 2014), with
turbulence maintained in the clump by protostellar outflows and/or by
accretion (Klessen \& Hennebelle 2010; Goldbaum et
al. 2011). Turbulent gas is expected to have a low rate of star
formation (Krumholz \& McKee 2005; Padoan et al. 2011), i.e.,
$\epsilon_{\rm ff}\sim0.02$. Such estimates are consistent with
observed protoclusters (e.g., Krumholz \& Tan 2007; Krumholz et
al. 2012; Da Rio et al. 2014). Since $\epsilon$ needs to be $\gtrsim
0.3$ to form bound star clusters, then, at least in such systems,
these estimates of $\epsilon_{\rm ff}$ imply $t_{\rm *cl,form}
\sim10\bar{t}_{\rm ff,cl}$.



The observational search for massive PSCs in protocluster clumps is
described in \S\ref{S:observation_PSC}, however, several points can be
considered here to help guide expectations. First, since massive stars
are rare, massive PSCs will also be rare objects. Most mass in a clump
will not be part of a massive PSC. If we define massive PSCs as having
$\gtrsim16\:M_\odot$, i.e., able to form $\gtrsim 8\:M_\odot$ stars
and if the PSCMF is described by a Salpeter (1955) power law of form
$dN/d{\rm log}M_c\propto M_c^{-\alpha}$ with $\alpha=1.35$ with lower
limit of $M_c=0.2$ or $1\:M_\odot$ (so that resulting stellar IMFs
approximately bracket the characteristics of the observed IMF) and
upper limit of $240\:M_\odot$, then the fraction of mass that is in
massive cores is 0.144 or 0.272, respectively. Thus a typical mass
fraction of cores that can form massive stars is $\simeq 0.2$. If the
fraction of the total clump mass that forms PSCs is $\simeq 0.5$,
i.e., so that total eventual star formation efficiency is $\simeq
0.25$, then only $\sim10\%$ of the total clump mass is expected to be
processed through massive PSCs. As discussed above, the protostellar
phase is expected to take $t_{*f}\simeq 1 \bar{t}_{\rm ff,cl}$ and
also $t_{*f}\simeq4.4\bar{t}_{\rm ff,c}$ (MT03). For a steady star
formation rate and, for simplicity considering closed box models
that take $10\bar{t}_{\rm ff,cl}$ to form, then at any given time the
observed massive protostellar core population will reflect only 10\%
of the total that ultimately forms and will only contain 1\% of the
initial clump mass. Similarly, if the same number of massive PSCs are
observed as massive protostellar cores, then this would reflect those
PSCs that are within $\simeq4.4\bar{t}_{\rm ff,c}$ before they form a
star (with $\bar{t}_{\rm ff,c}$ defined at this time). The
implications of observed PSC and protostellar core demographics are
described in \S\ref{S:observation_PSC}.


Competitive Accretion and Protostellar Collision models both predict
that massive protostars will be found near the centers of forming
clusters in regions of high (proto)stellar densities. However, Core
Accretion models may also predict a preference for massive cores to
form in denser regions near clump centers, e.g., if massive PSC
formation occurs via an agglomeration of smaller PSCs. More isolated
massive PSCs are also possible in Core Accretion models. There is a
general expectation that massive protostars forming in crowded regions
that suffer frequent tidal interactions with nearby stars will have
smaller accretion disks and more disturbed accretion geometries. For
example, the orientation of the disk and outflows would vary more in
Competitive Accretion than Core Accretion models. Strong accretion
variability, i.e., bursts, would also be expected to result from these
interactions. However, accretion bursts due to disk instabilities and,
more slowly, via infall variation in turbulent cores, are also possible
in Core Accretion models.

\section{Observational Studies of Massive Star Formation}

\subsection{The Search for Massive Pre-Stellar Cores}\label{S:observation_PSC}

\begin{figure}[t]
\vspace{-0.28cm}
\begin{center}
\includegraphics[width=5.3in]{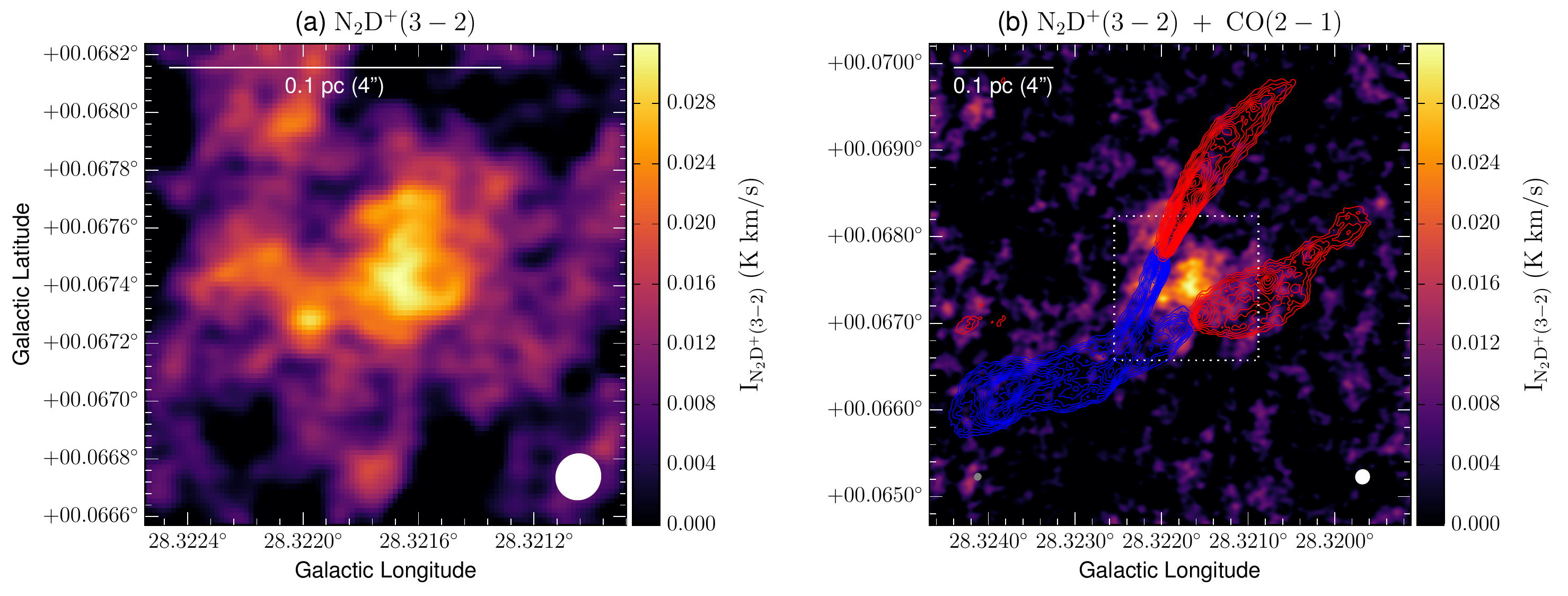} 
\vspace{-0.7cm}
 \caption{
(From Kong et al. 2017b) {\it ALMA} observations of integrated
   intensity of $\rm N_2D^+$(3-2) emission from the G028.37+00.07
   (C1-South) massive PSC candidate. The left panel shows a close-up
   view of the core, with the image smoothed to
   0.5 arcsec resolution. The right panel shows an expanded view of
   the region, now including high velocity CO(2-1) emission revealing
   two collimated bipolar outflows from nearby, but separate,
   protostellar sources.
}
\label{fig:n2dp}
\end{center}
\end{figure}

One of the best studied ``low-mass'' PSCs is L1544 (Caselli \&
Ceccarelli 2012). However, this core actually has $\sim8\:M_\odot$,
i.e., is quite massive, and its slow, subsonic infall
($\lesssim$10\% of free-fall), suggests $B$-fields play a significant
role in its dynamics (Keto et al. 2015).

Searches for more massive PSCs in higher $\Sigma_{\rm cl}$
environments have focused on IRDCs. Tan et al. (2013), following up a
sub-sample of the Butler \& Tan (2012) MIR extinction map peaks,
identified six cores via $\rm N_2D^+(3-2)$ emission. Follow-up
observations of the C1-S source identified protostellar outflows in
its vicinity (Tan et al. 2016), but analysis of the highest
resolution, highest sensitivity data (Kong et al. 2017b) indicates
that C1-S, as defined by its $\rm N_2D^+(3-2)$ emission, is spatially
and kinematically distinct from these sources and is thus a promising
massive PSC candidate (Fig.~\ref{fig:n2dp}). The mass estimate, based
on mm dust continuum emission, is about $50\:M_\odot$ inside a radius
of 0.045~pc, depending on assumed dust temperature (fiducial value of
10~K). This implies a mean density of $n_{\rm
  H}\simeq4\times10^6\:{\rm cm}^{-3}$. Note that due to systematically
cooler temperatures, PSCs like C1-S do not stand out as strong
continuum sources, especially compared to protostellar cores.

The velocity dispersion of C1-S is measured to be 0.28~km/s, which is
about 1/3 of the level expected from virial equilibrium of the
fiducial TCM core (eq.~\ref{eq:sigma}). For the core to be in virial
equilibrium would require stronger large-scale $B$-fields, i.e.,
$\sim3\:$mG, so that the Alfv\'en Mach number is about 0.2. Such
$B$-field strengths are similar to those predicted using the empirical
relation $B_{\rm median}\simeq 0.22 (n_{\rm H}/10^5\:{\rm
  cm}^{-3})^{0.65}\:$mG (for $n_{\rm H}>300\:{\rm cm}^{-3}$) (Crutcher
et al. 2010), to those observed on larger scales in some IRDCs (Pillai
et al. 2015) and to values inferred in some massive protostars
(\S\ref{S:obs_HMC}).

The deuteration fraction of C1-S was measured by Kong et al. (2016) to
be $D_{\rm frac}^{\rm N2D+}\sim0.15$ to 0.7. These values are similar
to the equilibrium values of the K15 astrochemical models for the
relevant physical conditions of the core. The timescales to reach this
level of deuteration are $\gtrsim10^5\:$yr, significantly longer than
the $\sim2\times 10^4\:$yr free-fall time. Most chemodynamical models
require relatively slow collapse compared to free-fall ($\alpha_{\rm
  ff}\lesssim 0.3$). Conversely, the example simulated core of Goodson
et al. (2016), which undergoes more rapid collapse
(Fig.~\ref{fig:goodson}), does not reach such high levels of $D_{\rm
  frac}^{\rm N2D+}$. Still, rapid collapse models can be made
compatible if the starting value of OPR$^{\rm H2}$ is very low (but
which itself may then require multiple free-fall times) or if ortho to
para $\rm H_2$ conversion rates are dramatically sped up compared to
gas phase estimates (Bovino et al. 2017). Observational constraints on
OPR$^{\rm H2}$ are needed to help break these degeneracies. Br\"uncken
et al. (2014) achieved this for the low-mass protostellar core IRAS
16293-2422 A/B via observations of ortho- and para-$\rm H_2D^+$,
estimating it has a chemical age of $>1$~Myr, i.e., $>10 t_{\rm
  ff}$. Similar studies are needed of more massive cores.


To increase the sample size of massive PSCs, Kong et al. (2017a)
searched 30 IRDC clumps for $\rm N_2D^+$(3-2) emission. Several
promising candidates were detected. Dynamical analysis of the 6
strongest sources was carried out. Together with the 6 cores analyzed
by Tan et al. (2013), overall this sample of 12 intermediate- and
high-mass PSC candidates have observed velocity dispersions that are
quite similar, within a factor of $\sim$0.8, compared to the fiducial
virial equilibrium value of eq.~(\ref{eq:sigma}).

Cyganowski et al. (2014) reported G11.920.61-MM2 as a massive PSC
candidate. However, the non-detection of any molecular lines from this
source is peculiar and makes it difficult to assess the reliability of
the structure, e.g., via a dynamical mass measurement. The Cygnus X
N53 MM2 core (Bontemps et al. 2010) and G11P6-SMA1 (Wang et al. 2014)
are other potential massive PSCs based on the absence of obvious
outflows (see also Motte et al. 2017).
Sanhueza et al. (2017) searched IRDC G028.23-00.19 for massive
PSCs. Given its current mass of 1,500$\:M_\odot$, if it were to form a
star cluster of $\sim500\:M_\odot$, then the median expected mass of
the most massive star would be about $26\:M_\odot$ (for Salpeter IMF
from 0.1 to $120\:M_\odot$), so a $\sim50\:M_\odot$ PSC is expected at
some stage in the clump. However, Sanhueza et al. find five cores with
masses up to only $\sim15\:M_\odot$, though these do appear to be
starless, i.e., lacking outflows. Such results may indicate that the
most massive core has not yet formed. Note, if cluster formation is
slow, then at any instant, only a small fraction $\sim\epsilon_{\rm
  ff}/\epsilon$ of the core population would be present.

Motte et al. (2007) and Russeil et al. (2010) (see also Motte et
al. 2017) estimated massive PSC and starless clump lifetimes as short
as $\lesssim 1$ to $3\times10^4\:$yr in Cygnus X and NGC6334/NGC6357,
by comparing to numbers of O to B3 stars and assigning a timescale of
a few Myr to these stars. In addition to the already mentioned
difficulty of identifying PSCs via dust continuum if they are
systematically colder than protostellar cores and the ambient clump
(e.g., Russeil et al. adopt a fiducial temperature of 20~K for their
mass estimates), another potential problem with this analysis is that
only core/clumps of $\geq40\:M_\odot$ and $\geq200\:M_\odot$ were
counted in Cygnus X and NGC6334/NGC6357, respectively. For example, in
the fiducial TCM, PSCs with masses $\sim16\:M_\odot$ are expected to
be able to produce $\sim8\:M_\odot$ stars, i.e., B3 stars on the zero
age main sequence.

\subsection{Massive Protostars, Accretion Disks, Outflows and Hot Molecular Cores}\label{S:obs_HMC}

Csengeri et al. (2017) studied mm dust continuum emission from 35
infrared quiet massive clumps, finding many examples of massive,
protostellar cores that show limited fragmentation: most regions are
dominated by just one or a few cores. The presence of strong
$B$-fields is a plausible explanation for the limited degree of
fragmentation in these sources, rather than, e.g., radiative
heating. On the other hand, Cyganowski et al. (2017) have argued that
there is a relatively high degree of fragmentation present in the
G11.92-0.61 region. If massive stars are forming in an unbiased way
within protoclusters, then, even in the context of Core Accretion
models, one expects that many lower-mass protostellar cores will also
be found in their vicinity.

Dynamically strong $B$-fields in massive protostellar cores have been
inferred by Girart et al. (2009) and Zhang et al. (2014) from
mm/sub-mm polarization observations. Vlemmings et al. (2010) measured
even stronger $B$-field strengths of $\sim20\:$mG via 6.7~GHz methanol
maser emission within $\sim1,000\:$AU of Cep A HW2.

Infall has been detected in 9 sources by Wyrowski et al. (2016),
though it can be difficult to tell if this is at the clump or core
scale. These authors find slow infall speeds: on average only
$\sim10\%$ of the free-fall speed. Processes that may slow infall
include support from $B$-fields and/or maintenance of clump turbulence
by outflows and accretion.

The search for and characterization of rotationally supported disks
remains challenging, which is not unexpected if diameters are
$\lesssim1,000$~AU, i.e., $\lesssim0.5$'' at 2~kpc. Since the review of
T14, there have been several claims of detection of such disks (e.g.,
Ilee et al. 2016; Beuther et al. 2017a), with these studies utilizing
emission of $\rm CH_3CN$. In the latter work, the authors achieve
130~AU resolution, find limited fragmentation on the core scale and
conclude the disk itself is also stable with respect to gravitational
instability.

Collimated molecular outflows are often a feature of massive
protostars (e.g., Beuther et al. 2002; Duarte-Cabral et al. 2013; see
also Fig.~\ref{fig:n2dp}, where the northern protostellar source is
estimated to have a core mass of $\sim30\:M_\odot$, Kong et
al. 2017b). On small scales relevant to outflow launching, Hirota et
al. (2017) have presented a high-resolution study of the closest
example of a massive protostar, Orion source I, finding evidence of
rotation near the base of the outflow, consistent with disk wind
models. During the later stages of formation and for the more massive
systems, the outflows are expected to become photoionized by the
protostar. Ionized, collimated outflows traced as radio continuum
``jets'' have been seen in some massive protostars (e.g., Gibb et
al. 2003; Guzm\'an et al. 2014), although the relative importance of
shock- versus photo-ionization remains to be established. Centimeter
continuum emission from ionized gas remains the most promising method
to identify the precise locations of massive protostars over a range
of evolutionary stages (e.g., Rosero et al. 2016), including detecting
the presence of multiplicity (e.g., Beuther et al. 2017b). A growing
sample of massive protostars, such as G35.20-0.74N, now have their IR
to mm SEDs well-characterized and fit to predictions of the TCM (Zhang
et al. 2013b; De Buizer et al. 2017). Elongation in the images from 10
to 40$\:{\rm \mu m}$ is expected along the outflow cavity and this
information helps to constrain the RT models. The goal of such studies
is to determine to what extent simple, symmetric protostellar models
can explain the observed dust continuum and, eventually, spectral line
morphologies. The presence of order and symmetry in core and outflow
features, especially when maintained over large scales, is not
expected in Competitive Accretion and Protostellar Collision
models. Core Accretion models may also exhibit spatial asymmetries,
e.g., due to low-order multiplicity resulting from disk fragmentation
and/or disk axis precession, as well as temporal variability, e.g.,
due to disk instabilities. Accretion bursts revealed by luminosity
variations have been reported by Caratti o Garatti et al. (2017) and
Hunter et al. (2017).

There are some examples of more disordered outflows, with the
larger-scale outflow from the Orion KL region, potentially driven by
source I, being a prime example (e.g., Bally et al. 2017). Dynamical
interaction among protostellar and young stellar sources, some of
which are now high proper motion runaway stars (e.g., Luhman et
al. 2017) seems likely to have played a role in triggering the
apparently ``explosive'' outflow, although the precise details of how
this has occurred remain debated (e.g., Bally \& Zinnecker 2005;
Chatterjee \& Tan 2012). It should be noted that the chemical
complexity of the Orion Hot Core (e.g., Schilke et al. 2001; Crockett
et al. 2010) is likely to have been affected by the strong shocks
resulting from this enhanced outflow activity. Another potential
example of an explosive outflow is the DR21 system (Zapata et
al. 2013), but such systems appear to be relatively rare in the
massive protostar population.

\section{Summary and Outlook}

Massive star formation involves many different complex physical and
chemical processes that need to be followed over vast ranges of
spatial and temporal scales. The initial conditions of the problem are
poorly constrained and often poorly defined. However, there is
progress driven by improving theoretical/computational modeling and
improving observational capabilities. Astrochemical modeling has
crucial roles to play in helping to carry out physical modeling, e.g.,
of ambipolar diffusion during the collapse of pre- and protostellar
cores, and for interpretation of observational signatures of the
various evolutionary phases of the massive star formation
process. However, given the large uncertainties present in both
physical and chemical models, great caution is needed when developing
and interpreting model results. Careful testing of predictions against
observations to then refine the models is essential.

\acknowledgements We thank Chris McKee for helpful comments on the
manuscript. We thank Maria Drozdovskaya, Matthew Goodson and Shuo
Kong for providing figures from their papers.



\end{document}